# Substrate-metal interface engineering enhances TaN/Ta thin film superconducting resonator performance

Moritz Singer
School of Computation, Information and Technology
Technical University of Munich
Garching, Germany
moritz.singer@tum.de

Harsh Gupta
School of Computation, Information and Technology
Technical University of Munich
Garching, Germany
harsh.gupta@tum.de

Benedikt Schoof
School of Computation, Information and Technology
Technical University of Munich
Garching, Germany
benedikt.schoof@tum.de

Elena Willinger
Electron Microscopy Facility
Technical University of Munich
Garching, Germany

Anton Orekhov
Electron Microscopy Facility
Technical University of Munich
Garching, Germany

Alexandra Schewski
Fraunhofer Institute for Electronic Microsystems and Solid State Technologies, Munich, Germany

Samuil Kostadinov
School of Computation, Information and Technology
Technical University of Munich
Garching, Germany

Philip C. Schneider
School of Computation, Information and Technology
Technical University of Munich
Garching, Germany

Marc Tornow
School of Computation, Information and Technology
Technical University of Munich
Garching, Germany
tornow@tum.de

***Abstract*— Tantalum has been demonstrated as a promising material for superconducting qubits. However, comparatively little attention has been given to its nitrides. Tantalum nitride exhibits a range of stoichiometries, resulting in a variety of material properties, including both superconducting and non-superconducting phases. Owing to this versatility, tantalum nitrides can serve multiple purposes in superconducting qubits: as seed layers for alpha-Ta growth, as a superconducting base material and as a non-superconducting barrier in the Josephson junction. In this study, we explore the performance of superconducting TaN and Ta thin film combinations on silicon substrates in terms of internal quality factor $Q_i$. We find that standalone TaN films exhibit $Q_i$ values of about $1.5\times10^5$ at 100mK in the single-photon regime. Surprisingly, a resonator made from Ta grown on a few-nanometers-thick TaN seed layer yields largely the same performance. However, adding an additional, few-nanometers-thick Ta buffer layer between the Si substrate and this TaN seed layer enhances $Q_i$ significantly up to $5.9\times10^5$. Supporting transmission electron microscopy measurements reveal structural disorder at the TaN-Si interface, while this interfacial modification is suppressed when the Ta buffer layer is introduced. The observed improvement in resonator performance is consistent with a reduction of interface-related two-level system losses and strongly supports the hypothesis that controlling the substrate-metal interface is pivotal for the performance of superconducting qubit circuitry.**

**Keywords—Tantalum nitride, thin films, superconductivity, quantum computing, coplanar waveguide resonators, metal-substrate interface**

## I. INTRODUCTION

Qubits can be realized by building so-called artificial atoms in circuit quantum electrodynamics (cQED). The main required building blocks are coplanar waveguide (CPW) resonators, capacitances and Josephson junctions (JJs) [1], [2]. The CPW resonators are used for initialization, control and readout of the qubit and are made of a superconducting material. The JJ provides the non-linear inductance that is needed to make the potential of the LC-oscillator of the circuit anharmonic, enabling individual addressing of its transitions [3], [4]. The CPW resonators are routinely used to benchmark the performance of superconducting thin films in the single-photon regime where two-level system (TLS) losses are dominant [5]. For the superconductor in qubits, various materials have already been explored [6], [7]. Amongst others, elements like niobium (Nb) [8], [9], [10], tantalum (Ta) [11], [12], [13] and aluminum (Al) [14], [15] , as well as nitrides like niobium nitride (NbN) [16] and titanium nitride (TiN) [17], [18] have been used to fabricate superconducting quantum circuits [6], [7]. However, so far comparatively little attention has been given to the nitrides of tantalum. Tantalum nitride can exhibit a range of stoichiometries [19], [20], including superconducting (delta-TaN) [21] as well as non-superconducting $Ta_3N_5$ and $Ta_5N_6$ [22]. This versatility enables the fabrication of superconducting qubit components exclusively from different forms of tantalum nitrides. TaN as a barrier material in JJs has already been demonstrated [23]. Superconducting CPW resonators made from TaN, have not been thoroughly investigated though. In this study, we optimized nitrogen ($N_2$) to argon (Ar) flow rates and sputtering pressure to deposit tantalum nitride thin films on Si (100) substrates. We use transmission electron microscopy (TEM) to get high-resolution cross-sectional images of the materials and to confirm the thicknesses of the layers. We further analyzed the stacks with electron energy loss spectroscopy (EELS) to investigate the contents and the interfaces. The films were characterized in terms of surface roughness, crystallography, resistivity, and superconducting transition. CPW resonators were fabricated from a selection of the thin films that turn superconducting and the internal quality factors ($Q_i$) were measured at 100 mK. Furthermore, the same TaN films were used as 5 nm thick seed layers, on which 200 nm Ta was deposited. Ta sputtered on TaN grows in the alpha-phase, as shown in a previous study [24]. These stacks were also characterized with the same methods, and the RF-performances are compared. As $Q_i$ of the Ta on the TaN seed layers is similar to the thin films with TaN only, we further investigate the influence of the substrate-to-metal interface on the $Q_i$. Therefore, we fabricated the stack with 200 nm of Ta on the 5 nm TaN seed layer this time introducing an additional 10 nm Ta layer between the Si substrate and TaN to modify the interface. We compare the results from these stacks in terms of the RF-performance of the CPW resonators.

## II. EXPERIMENTAL

### A. Deposition of tantalum (nitride) thin films

The thin films in this study were sputter-deposited on high-ohmic (>10 kΩ·cm) silicon (100) substrates in an RF-magnetron sputter tool (Alcatel A450). The silicon was diced into 10 mm x 6 mm chips from 4-inch wafers. The chips were chemically cleaned by ultrasonication in acetone and isopropanol for three minutes respectively and subsequently dipped into piranha solution for 10 minutes (3:1 $H_2SO_4$:$H_2O_2$, *Caution: piranha solution is a highly corrosive oxidizing mixture that reacts violently with organics; special care/training is mandatory*). After this step, the chips were ultrasonicated in isopropanol again for 1 minute and dipped into buffered oxide etch (BOE, 7:1 HF:$NH_4F$=12.5:87.5, *Caution: hydrofluoric acid (HF(aq)) is very hazardous to health; special care/training is mandatory*) for 60 seconds. The latter step is done to remove the native silicon oxide. After the chemical cleaning, the substrates were put for 10 minutes on a hotplate which was heated to 200°C, to diffuse out hydrogen on the surface that is left over from the rinsing. After this, the cleaned substrates were mounted into the chamber of the aforementioned sputtering tool. The sputter chamber's base pressure, at which the sputtering took place, was ~$10^{-7}$ mbar. Reactively sputtered tantalum nitride was deposited by introducing nitrogen to the sputter chamber, alongside the process gas argon. The flow rate of Ar was 20 sccm, while the $N_2$ flow rate was 2 sccm at a constant gas pressure of 18 μbar. The target-to-substrate distance was 7 cm and the material was deposited at room temperature at a power of 200 W. In case of Ta sputtering, the same process parameters were used, with the only difference that there was no $N_2$ gas flow. For stacks of multiple materials, the substrates stayed in the vacuum after the first deposition was finished. The subsequent layers were deposited without breaking the vacuum to avoid oxidation of the material. The thickness of the sputtered layers was determined by sputtering on substrates with a patterned photoresist. After the deposition, a lift-off process was performed and the height of the remaining structures relative to the Si surface was measured with a profilometer for thicknesses >10 nm and with an atomic force microscope (AFM, Bruker (Veeco) Dimension V-AFM) for layers <10 nm. The sputtering time was optimized to obtain either 200 nm thin films of TaN only, or a 5 nm thick TaN seed layer, followed by 200 nm of Ta, or - for a third stack - first 10 nm of Ta, followed by 5 nm of TaN and finally 200 nm of Ta. A schematic sketch of the three different thin film stacks is shown in Fig. 1 (a).

### B. Thin film characterization methods

The surface morphology of the planar tantalum thin films was investigated with the before-mentioned AFM in tapping mode. An area of 5 μm x 5 μm was scanned with a rate of 1 Hz and 512 data points per scan line, and the root mean square roughness *Sq* was determined from the obtained pictures. To investigate the crystallographic structure of the material, planar chips were measured by the grazing incidence X-ray diffraction (GI-XRD) method. In the tool that was used (Panalytical Empyrean), measurements were done with a fixed angle of 0.5 degree of the incidence beam and a range of 20-90 theta angles of the detector beam. A dual-beam focused ion beam (FIB) instrument (Zeiss Crossbeam 550) was used to cut a lamella out of the thin film stacks. A TEM tool (Jeol JEM-F200) was used to take high resolution images of the cross-sections from which the disorder in the crystal lattice at the interface is assessed. The same cross-sections were analyzed by EELS to qualitatively determine the depth profile of the constituent elements. To measure the resistive and superconducting properties, planar films were patterned into test structures comprising of a bar with pads attached to its sides. With this configuration, a four-point measurement was done by contacting the bar on both ends and forcing a current through, while measuring the voltage drop between two pads along the side. As the geometry of the bar and the distance of the pads is known by design (300 μm width, 3600 μm length, two pads on each side 1500 μm apart) and the thickness of the sputtered thin films is also known, the resistivity of the material was calculated with Ohm's law. For critical temperature measurements, the chips were bonded to a PCB and mounted into an adiabatic demagnetization cryostat (kiutra) and cooled down to sub-K temperatures. While measuring continuously in the aforementioned four-point configuration, the temperature was swept from 200 mK to 7 K and back down. The temperature sweep was performed in both directions to check if the thermalization of the sample has a delay, which would manifest itself as an offset between the two superconducting transitions. This was not present in our data, which confirms that the chips thermalize without any delay and indeed have the temperature that is measured by the sensor on the sample stage.

### C. Fabrication and measurements of resonators

The RF-performance of the thin films was determined with CPW resonators in a hanger configuration. The resonators were designed as $\lambda/4$ resonators with a center conductor line width of 10 μm and a gap width of 6 μm, resulting in an impedance of approximately 50 Ω. Twelve resonators in the frequency range 4 to 8 GHz and capacitively coupled to a common feed line were patterned by photolithography and RIE etching with $CF_4$ gas. The etch depth was optimized to over-etch the thin films 10-20 nm into the silicon substrates. The etching was followed by photoresist removal by ultrasonication in acetone and isopropanol and the samples were again immersed in BOE for 10 minutes. After these steps, the pads at the ends of the feedline were bonded to a PCB, which was mounted into a custom-made copper box, which connects via two SMAs to the wiring through the cryostat. $S_{21}$-measurements were done with a vector network analyzer P5002B (Keysight) by sending RF-signals into the chip and detecting the transmitted signal. From a coarse frequency sweep over 4-8 GHz, the resonance peaks of the CPW resonators were detected and fine sweeps around the peaks were done. After this, each resonance frequency was measured individually with input powers ranging from -90 dBm to -160 dBm in -10 dBm steps. A circle fit was done to obtain the loaded quality factor $Q_l$, as well as the coupling and the internal quality factors, $Q_c$ and $Q_i$, respectively, of each resonator [25]. The resonators were designed to operate in the over-coupled regime, which means that $Q_i > Q_c$.

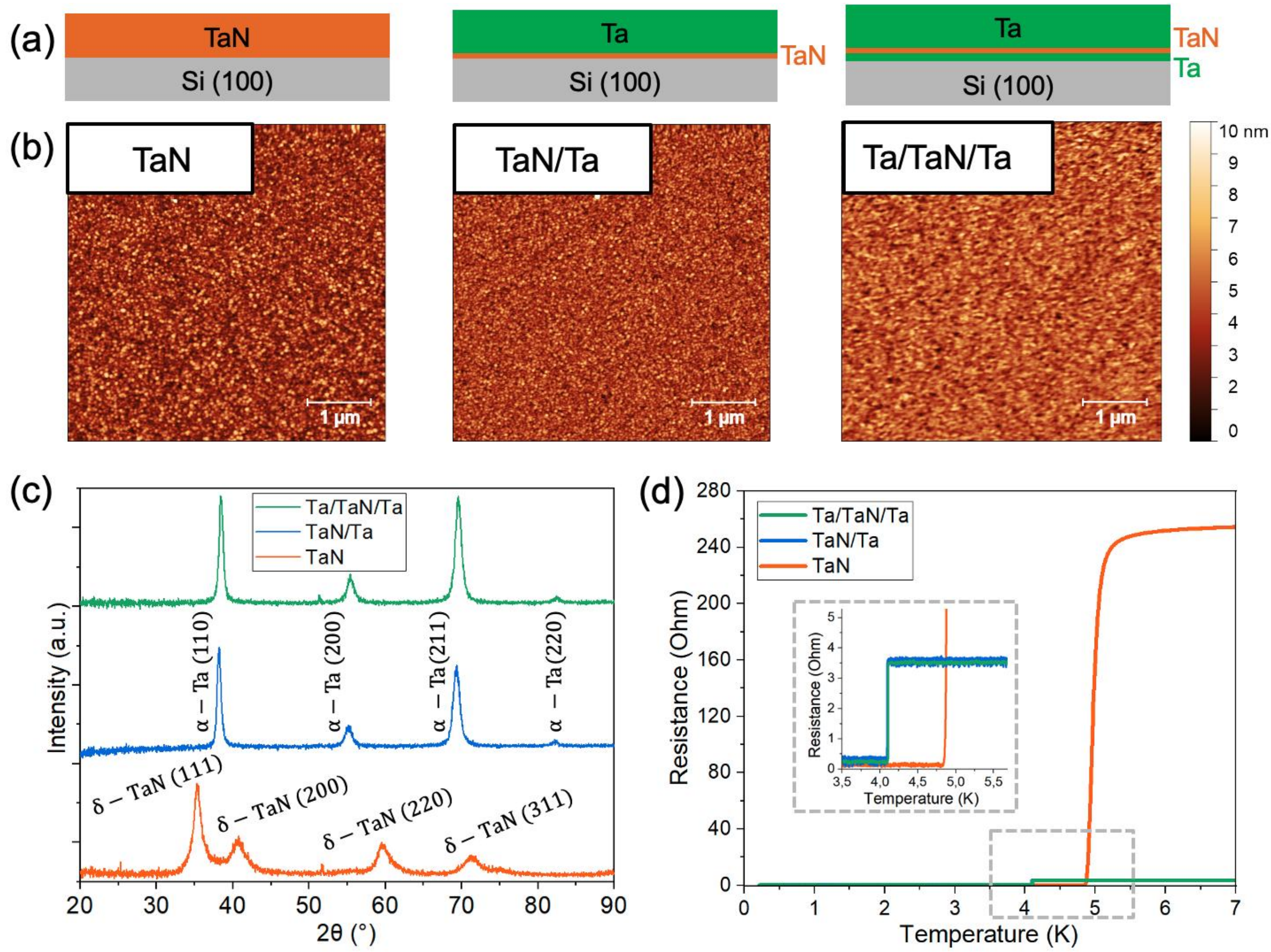

*Figure 1: (a) Schematic sketches of the sputtered material stacks: 200 nm thick of TaN, as well as 5 nm TaN followed by 200 nm of Ta and 10 nm of Ta, followed by 5 nm of TaN and 200 nm of Ta were deposited on Si (100) substrates. (b) AFM pictures of the three deposited thin film stack surfaces. The surface roughness is in the range of 0.4-0.9 nm for all three films. (c) GI-XRD spectra of the thin films, showing exclusively alpha-peaks in Ta/TaN and Ta/TaN/Ta. The TaN spectrum shows peak associated with cubic delta-TaN. (d) Critical temperature measurements, where a superconducting transition was observed for all stacks. Ta on TaN and on TaN/Ta turns superconducting at ~4.1 K, while the TaN thin film turns superconducting at ~5 K.*

## III. Results and Discussion

### A. Surface, crystallographic and electrical characterization

The AFM pictures of the material stack surfaces are shown in Fig. 1 (b). The surface of all thin films is relatively smooth with RMS roughness values of 0.9 nm for the TaN-only thin film, 0.6 nm for the Ta on top of the TaN seed layer and 0.4 nm for the Ta on top of the 5 nm TaN and Ta layers. The GI-XRD spectra of the thin films are shown in Fig. 1 (c). For the Ta/TaN/Ta as well as the TaN/Ta stack, four peaks are present in the spectrum, all of which can be attributed to the four alpha-Ta peak planes (110, 200, 211, 220). This indicates that alpha-Ta has grown on top of the TaN seed layer and on top of the TaN seed layer with the Ta buffer layer underneath. The TaN spectrum, as expected, looks substantially different from the Ta spectrum. In this case, the peaks can be attributed to cubic delta-TaN [26], [27].

The critical temperature measurements are shown in Fig. 1 (d). All three stacks show a superconducting transition, the stacks with Ta on TaN and on TaN/Ta at ~4.1 K. The TaN only thin film turns superconducting at ~5 K, which is in the range of previously reported values [28], [29]. The residual resistance of stacks with 200 nm alpha-Ta is much lower than the one for TaN, probably due to the lower room temperature resistivity of 25 µΩ·cm compared to 810 µΩ·cm of TaN. The residual resistance ratio (RRR) for the stacks with Ta are $RRR_{TaN/Ta} = 2.36$ and $RRR_{Ta/TaN/Ta} = 2.41$. The TaN thin film shows a different behavior as the resistance is lower at room temperature and the $RRR_{TaN}$ value is therefore smaller than one (0.76) in this case, which indicates high disorder in the thin film, which is a known effect in sputtered TaN on silicon [30].

### B. RF-measurements of CPW resonators

The data of the RF-measurements of CPW resonators are shown in Fig 2. As expected, the $Q_i$ increases with power, as two level system losses (TLS) are expected to be the dominating losses in the low power regime and begin to saturate with increasing power. It is also visible that the TaN and TaN/Ta stacks perform similarly, while the Ta/TaN/Ta shows significantly higher $Q_i$ values. The difference is evident across the full power range and becomes more pronounced in the low-power limit, consistent with TLS losses setting the dominant performance constraint. Further, TaN and TaN/Ta show a distinct plateau of $Q_i$, while the $Q_i$ of Ta/TaN/Ta levels off at lower powers. To extract the intrinsic quality factor corresponding to TLS loss in the low-power regime $Q_{TLS,0}$ and the critical photon number $n_c$, the data was fitted to the following TLS-model [31]:

$$\delta_{TLS} = \frac{F\delta_{TLS}^{0} tanh\left(\frac{\hbar\omega}{2k_B T}\right)}{\sqrt{1+\frac{n}{n_c}}} \qquad (1)$$

where $F$ is the filling factor, $\delta_{TLS}^{0}$ is the intrinsic TLS loss ($F\delta_{TLS}^{0} = 1/Q_{TLS,0}$) and $n$ is the mean number of photons circulating in the resonator. For the photon number calculation [32], [33], the quality factors $Q_l$ and $Q_c$ are obtained from the circle fits. As per design, $Q_c$ is in the range of $1\times10^5$ [34]. The fits for each data set are represented by the dashed lines in Fig 2.

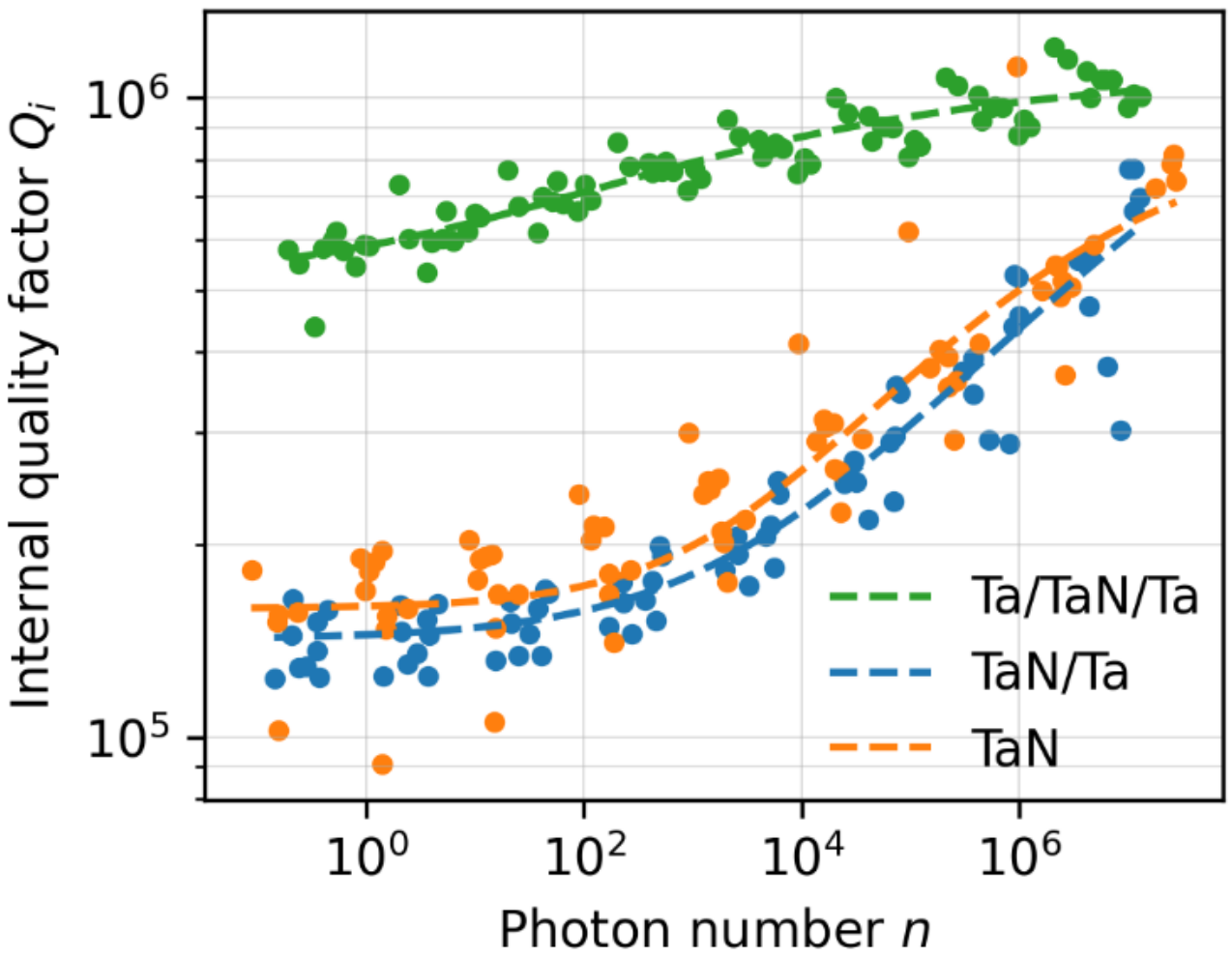

*Figure 2: Internal quality factors of thin film CPW resonators measured at 100 mK, as function of the mean photon number n. In the single-photon regime, TaN and Ta on TaN have similar Qi of $Qi_{TaN}$ = $1.53\times10^5$ and $Qi_{TaN/Ta}$ = $1.41\times10^5$. In contrast, the stack with the added 10 nm Ta has a significantly higher $Qi_{Ta/TaN/Ta}$ = $5.88\times10^5$. In the high-power regime, the Ta/TaN/Ta stack still has higher Qi, but the difference to the other thin films is much smaller than in the single-photon regime. Circle symbols: measured data points.. Dashed lines: numerical fits to eq. 1.*

It is striking that the 200 nm thick Ta on top of the 5 nm TaN seed layer performs similarly as the 200 nm TaN thin film. In the single photon regime, both have a similar $Qi$ value ($Qi_{TaN}$ = $1.53\times10^5$ and $Qi_{TaN/Ta}$ = $1.41\times10^5$), but also in the high-power regime they perform approximately the same with values approaching $8\times10^5$. The stack with the 10 nm Ta between the TaN seed layer and Si, however, performs significantly different. While its high-power $Q_i$ is only slightly higher with just above one million, in the single-photon regime it has a much higher $Qi_{Ta/TaN/Ta}$ of $5.88\times10^5$, compared to the other two stacks. The same trend is visible in the TLS-limited internal quality factor extracted from the fits. Here, the value for TaN ($Q_{TLS,0}$ = $1.64\times10^5$) is in the same range as for TaN/Ta ($Q_{TLS,0}$ = $1.36\times10^5$) while the Ta/TaN/Ta value ($Q_{TLS,0}$ = $9.06\times10^5$) is significantly higher. In a previous study, pure alpha-Ta thin films were grown under the same conditions directly on Si by high-temperature sputtering. The $Q_i$ values in this case reached >$10^6$ in the single-photon regime [35]. The critical photon number of the TaN ($1.71\times10^3$) and TaN/Ta ($2.58\times10^3$) stack is also similar, while it is far smaller for Ta/TaN/Ta (4.08), where TLSs begin to saturate at very low powers.

These observations strongly suggest that the TaN to Si interface is the limiting factor in the TaN and TaN/Ta stacks. Our initial characterization described above confirmed that these two thin-film stacks are very much different, as anticipated for one comprising exclusively TaN while the other one mostly consisting of alpha-Ta, exhibiting for example different electrical properties. They have different metal-to-air interfaces (TaN to air and almost exclusively alpha-Ta to air), but they have the same metal-to-substrate interface (TaN to Si). Adding the 10 nm Ta buffer layer, changes this interface and results in an almost 4-fold increase in $Q_i$ in the single-photon regime. Further, the Ta buffer layer results in a much lower critical photon number. This suggests that the TLSs are strongly coupled to the electrical field and saturate with less power [36], [37]. In CPW resonators, regions of high electric field concentration are located at the sidewalls and edges. This lets us assume that in the stack with the buffer layer, we see $Q_i$ mostly influenced by surface TLSs at the metal-to-air interface. Since the Ta buffer layer changes only the metal-to-substrate interface, we hypothesize that the dominating TLS are concentrated at the TaN to Si interface for the other two sample types. The substrate-to-metal interface can be the limiting loss channel, as has been demonstrated before [38]. For TaN and TaN/Ta, a higher critical photon number is observed, meaning that dominant TLSs are not as strongly coupled and saturate only at higher powers.

### C. Interface investigation by TEM

To further investigate the origin of the observed differences in resonator performance, the interface of the TaN/Ta (and Ta/TaN/Ta) stacks to Si was analyzed by TEM together with EELS. While TEM provides structural information, EELS was used to qualitatively map the elemental distribution across the interfaces. Cross-sectional TEM images of the TaN/Ta stack are shown in Fig. 3 (a). The images confirm the expected layer structure, consisting of Ta deposited on a TaN seed layer with a thickness of approximately 5 nm. The corresponding EELS maps further confirm the presence of a nitrogen-containing interlayer at the Si interface, consistent with the TaN seed layer. In the TaN/Ta stack, the nitrogen signal extends directly to the substrate interface, indicating that nitrogen is present at the Si-facing boundary. High-resolution TEM images of the TaN-Si interface reveal a region with structural disorder and lattice defects in the interface region. The disorder could originate from energetic particle bombardment during reactive sputtering, atomic peening or from chemical reactions at the surface during the early stages of TaN growth. In contrast, the EELS analysis of the Ta/TaN/Ta stack shows the nitrogen-containing region separated from the silicon by the inserted Ta buffer layer. This confirms that the Ta interlayer prevents direct contact between the reactively sputtered TaN layer and the Si substrate. The TEM images of the Ta/TaN/Ta stack are shown in Fig. 3 (b). Again, the expected layers of Ta on approximately 5 nm of TaN on 10 nm of the Ta buffer layer are visible. However, in contrast to the TaN/Ta stack, at the interface and in the Si no structural disorder or clusters of defects can be seen, as was the case before.

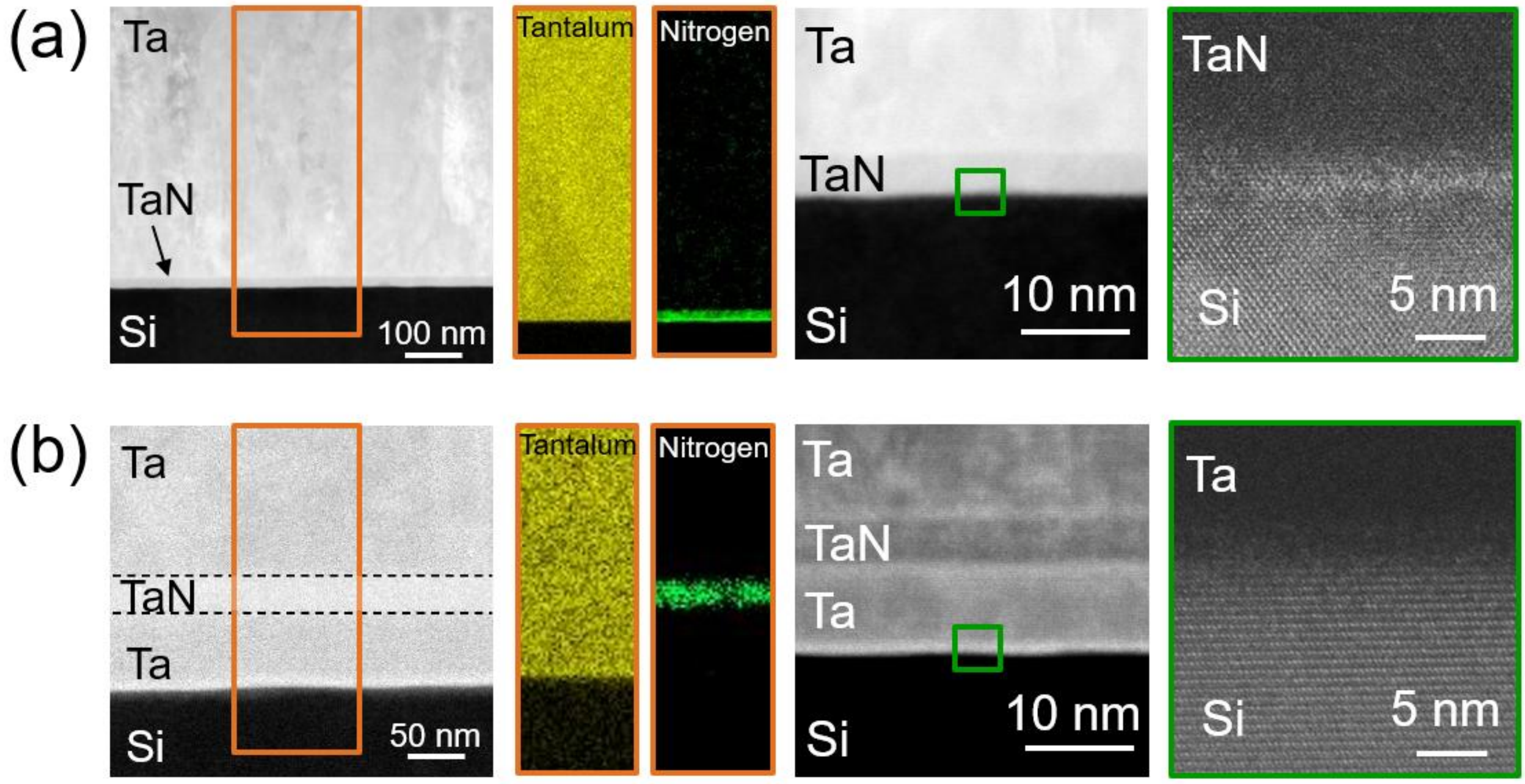

*Figure 3: (a) TEM pictures and EELS analysis of the TaN/Ta stack, confirming the expected layer stack of Ta on 5 nm of TaN on Si. In the high-resolution TEM images in the right, irregular lighter and darker regions next to the interface can be seen, indicating dislocations and disorder in the Si lattice. (b) TEM and EELS analysis for the Ta/TaN/Ta stack, where again Ta is deposited on 5 nm of TaN, but a 10 nm thick Ta buffer layer between TaN and Si is introduced. For this stack, no structural disorder or defects in the Si lattice are visible and the interface to the Si looks undistorted.*

From Fig. 3 it is evident that structural disorder is present in the TaN/Ta stack at the metal-substrate interface. As outlined earlier, this stack also exhibits significantly lower $Q_i$ in the single-photon regime. In comparison to that, the stack containing the Ta buffer layer shows no structural disorder and shows an almost four-fold higher value of $Q_i$. While this correlation does not provide direct proof of the microscopic loss mechanism, it is consistent with the hypothesis that defects at the TaN-Si interface, associated with reactive sputtering with nitrogen, are a major source of TLS losses. Taken together, the structural and compositional analysis suggests that the presence of nitrogen at the silicon interface is most likely introducing performance limiting losses, which can be suppressed by introducing a thin Ta buffer layer prior to the reactive TaN deposition. The corresponding improvement in resonator quality factor indicates that controlling the chemical composition of the metal–substrate interface is an important factor for reducing TLS-induced losses in superconducting devices.

## IV. CONCLUSION

In this work, we investigated reactively sputtered tantalum nitride thin films as a standalone material and in combination with Ta towards their suitability for superconducting microwave circuits. TaN films were characterized in terms of surface roughness, crystallinity and electrical properties. CPW resonators fabricated from these films were measured at 100 mK to evaluate their microwave performance in the single-photon regime. The measurements show that superconducting TaN thin films exhibit internal quality factors on the order of $Q_i = 1.5\times10^5$. When Ta is deposited on a TaN seed layer, the resulting stack shows a comparable microwave performance despite the different electrical properties of the two materials. This strongly indicates that the dominant loss mechanism in these structures is not determined by the bulk properties of the superconducting film but rather by the metal-to-substrate interface. A substantial performance gain is achieved by inserting a thin Ta buffer layer beneath the TaN seed layer. This modification increases the $Q_i$ to $5.9\times10^5$, representing an almost fourfold improvement relative to the other stacks examined. Structural and compositional analysis by TEM and EELS shows that stacks without the buffer layer exhibit disorder at the TaN-Si interface and nitrogen extending directly to the substrate boundary. Insertion of the Ta buffer layer leads to an undisturbed interface lattice structure and separates the nitrogen-containing TaN layer from the silicon. This correlation suggests that nitrogen-related interfacial defects form a dominant source of TLS that limits resonator performance. Preventing nitrogen incorporation at the metal–substrate interface may lead to a substantial improvement in the internal quality factor. The findings highlight the importance of interface engineering when integrating transition-metal nitrides into superconducting quantum circuits.

## Acknowledgment

We acknowledge funding through the Munich Quantum Valley (MQV) project by the Free State of Bavaria, Germany, and the Munich Quantum Valley Quantum Computer Demonstrators - Superconducting Qubits (MUNIQC-SC) project (grant no. 13N16188), funded by the Federal Ministry of Research, Technology, and Space, Germany. We would like to thank our colleagues from the TUM Catalysis Research Center (CRC) for giving us access to XRD-measurement equipment, the TUM Center for Nanotechnology and Nanomaterials (ZNN) for shared use of the clean room facilities, the technical staff working at TUM ZEITlab, and the colleagues at the Walther-Meissner-Institute for the initial resonator layout and copper box designs.